\documentclass[review,3p,times]{elsarticle}

\usepackage{amssymb,amsmath,amsfonts,mathrsfs,bm,gensymb}
\usepackage{graphicx,float,subfig}
\graphicspath{ {./figures/} }

\usepackage{geometry}
\usepackage{hyperref}
\usepackage{color}
\usepackage{soul}

\DeclareMathOperator{\sgn}{sgn}
\hypersetup{
    colorlinks=true,
    linkcolor=black,
    citecolor=black,
    filecolor=black,
    urlcolor=black,
}

\makeatother

\begin{document}

\begin{frontmatter}

\title{Cloaking strategy for Love waves}

\author[add1]{Z. Chatzopoulos}
\author[add1]{A. Palermo \corref{corr1}}
\ead{antonio.palermo6@unibo.it}
\author[add2]{S. Guenneau}
\author[add1]{A. Marzani\corref{corr1}}
\ead{alessandro.marzani@unibo.it}
\cortext[corr1]{Corresponding authors}

\address[add1]{Department of Civil, Chemical, Environmental and Materials Engineering, University of Bologna, 40136 Bologna, Italy}
\address[add2]{UMI 2004 Abraham de Moivre-CNRS, Imperial College, London SW7 2AZ, United Kingdom}

\begin{abstract}

Love waves are antiplane elastic waves which propagate along the surface of a heterogeneous medium. Under time-harmonic regime, they are governed by a scalar equation of the Helmholtz type. We exploit the invariance of this governing equation under an in-plane arbitrary coordinate transformation to design broadband cloaks for surface defects.
In particular, we apply transformation elastodynamics to determine the anisotropic, position dependent, mechanical properties of ideal cloaks able to hide triangular and parabolic-shaped defects. Dispersion analysis and time-harmonic numerical simulations are employed to validate the proposed strategy.
Next, we utilize layered monoclinic materials, with homogenized properties matching those of ideal cloaks, to design feasible cloaks. The performance of the layered cloaks is validated via time-harmonic numerical simulations which show a significant reduction of the defect-generated scattered fields.
\end{abstract}


\end{frontmatter} 
\section{Introduction}
\label{sec:intro}

The prospect of rerouting the propagation of elastic waves around an object and isolate it from unwanted mechanical vibrations has fuelled the research interest towards the realization of elastic cloaking devices. A cloak is a region of a medium surrounding a defect, or a target object, purposely designed to make it scattering free from particular waves. 
Among the approaches existing to design a  cloak, transformation elastodynamics has been extensively employed in recent works as it allows for the direct derivation of passive cloaks material properties ~\cite{Kadic_2013}. In brief, transformation elastodynamics consists in a coordinate transformation to map (or stretch) a spatial region embedding a defect, the cloak, to a region as if no obstacle were there. In general, this coordinate transformation turns a homogeneous and isotropic material into an inhomogeneous and anisotropic material, posing some challenges for the practical realization of a cloak. 

A necessary mathematical prerequisite for material-parameter transformation to work is that the underlying equations must be invariant with respect to general coordinate transformations. Milton et al.~\cite{Milton_2006} showed that, in the general case of an elastic medium, the equations of motion are not form invariant under an arbitrary transformation, leading to non-scalar density and additional third-order elastic tensors. Also, the choice of the map linking the displacements to their transformed counterpart is proven vital, since the symmetry of the stress tensor is dependent upon it. Brun et al.~\cite{Brun2009} realized that choosing an identity map linking these displacements makes the Navier equation transformation invariant, but breaks the minor symmetries of the transformed elasticity tensor.
Norris et al.~\cite{NORRIS2011} considered multiple types of map (constant, identity, etc), that resulted in having materials breaking the stress symmetry.
In these cases, the resulting material parameters are impractical and very difficult to handle. An ample number of ideas tackled those situations, including the adoption of the so-called Cosserat or micropolar cloaks with chiral behaviour. Chiral materials allow to obtain the required non-symmetric stress tensor, by enforcing an additional rotational degree of freedom in the form of a body torque~\cite{NASSAR2019a, NASSAR2019b} or rotational resonance~\cite{ZHANG2020103796}. Similarly, Achaoui et al.~\cite{Achaoui2019} considered a Willis medium as background material and, exploiting its form invariance~\cite{Milton_2006}, developed a Cosserat-Willis type cloak. 
Other strategies include the use of pentamode materials to achieve the required effective properties, although their mechanical instability is a considerable impedance factor for their practical manufacture \cite{Kadic2012}. 

Nonetheless, there are configurations where the governing equation retains its form, such as thin-plate geometries. In this context, Farhat et al. \cite{Farhat2009} proposed a cloak to control bending waves in isotropic heterogeneous thin plates for a wide range of frequencies. Subsequently, Stenger et al.~\cite{Stegner2012} building upon this theory, provided an experimental proof by fabricating an elastic cloak using homogeneous and locally isotropic rings. 
Further experiments by Misseroni et al. \cite{misseroni2016cymatics,misseroni2019omnidirectional} showed the possibility of cloaking flexural vibrations in structured plates. However, control of flexural pulse dynamics remains a further challenge \cite{pomot2019form,tang2021pulse}.
We remark that most of the above literature concerns passive cloaking designs, since it exploits the concept of geometric transformation, in the tracks of what  earlier proposed for electromagnetic waves. Some alternative path to cloaking using active sources can reduce the plate design complexity in the time harmonic regime \cite{o2016active}. This active cloaking should work equally well for acoustic and transverse elastic waves \cite{vasquez2009active,vasquez2013transformation}.

Besides the rich literature on passive and active cloaking for elastic waves, applications  for surface waves are still rare. The most relevant studies, to the best of our knowledge, are those that exploit micropolar materials  \cite{khlopotin2015} or near-cloaking techniques \cite{quadrelli2021} to hide surface defects from Rayleigh waves.
In this work, we explore the possibility of cloaking surface defects from waves of the Love type, a scenario that has not been considered yet. Love waves are antiplane shear waves propagating at the free surface of a heterogeneous medium with a stiffness profile increasing with depth.

We note that some control of Love waves has been achieved using locally resonant elements atop the air-soil interface \cite{palermo2018control,maurel2018conversion}, but these works do not make use of the concept of artificial anisotropy, as required by geometric (non conformal) transforms. 
Here instead, by exploiting the invariant form of antiplane shear waves, we apply transformation elastodynamics to design cloaks for surface defects. In particular, we focus our attention to carpet pinched cloaks able to hide a defect placed at the surface of the domain.

First, we apply transformation elastodynamics to normally dispersive profiles, i.e. with increasing shear speed for increasing depth of the medium, to design ideal cloaks. We describe the proposed mapping and provide the space dependent anisotropic properties of the mapped material. We prove, via numerical finite element simulations, the validity of the proposed mapping, showing both the invariance of the dispersive properties of Love waves and the near zero scattering for configurations with  cloaked defects.  

Next, for a surface triangular defect, we apply a straightforward homogenization technique to map the material properties of the ideal cloak onto a monoclinic double-material unit cell. The unit cell allows the realization of a layered cloak with a feasible microstructure.
We confirm the validity of the adopted homogenization step by comparing the performances of the layered cloak with those of the ideal cloak. Finally, we discuss potential and limitations of the proposed study.

\section{Statement of the problem}
We propose an approach to cloaking of Love waves that involves a coordinate stretch in the medium vertical plane, thus affecting  the surface wave trajectory along the same plane, see Fig.~\ref{Figure 1}. This approach is in stark contrast to  \cite{palermo2018control} where the wave control was achieved in the horizontal plane.

\subsection{Antiplane surface waves in vertically heterogeneous media} 

\begin{figure}
\centering
\includegraphics[scale=0.70]{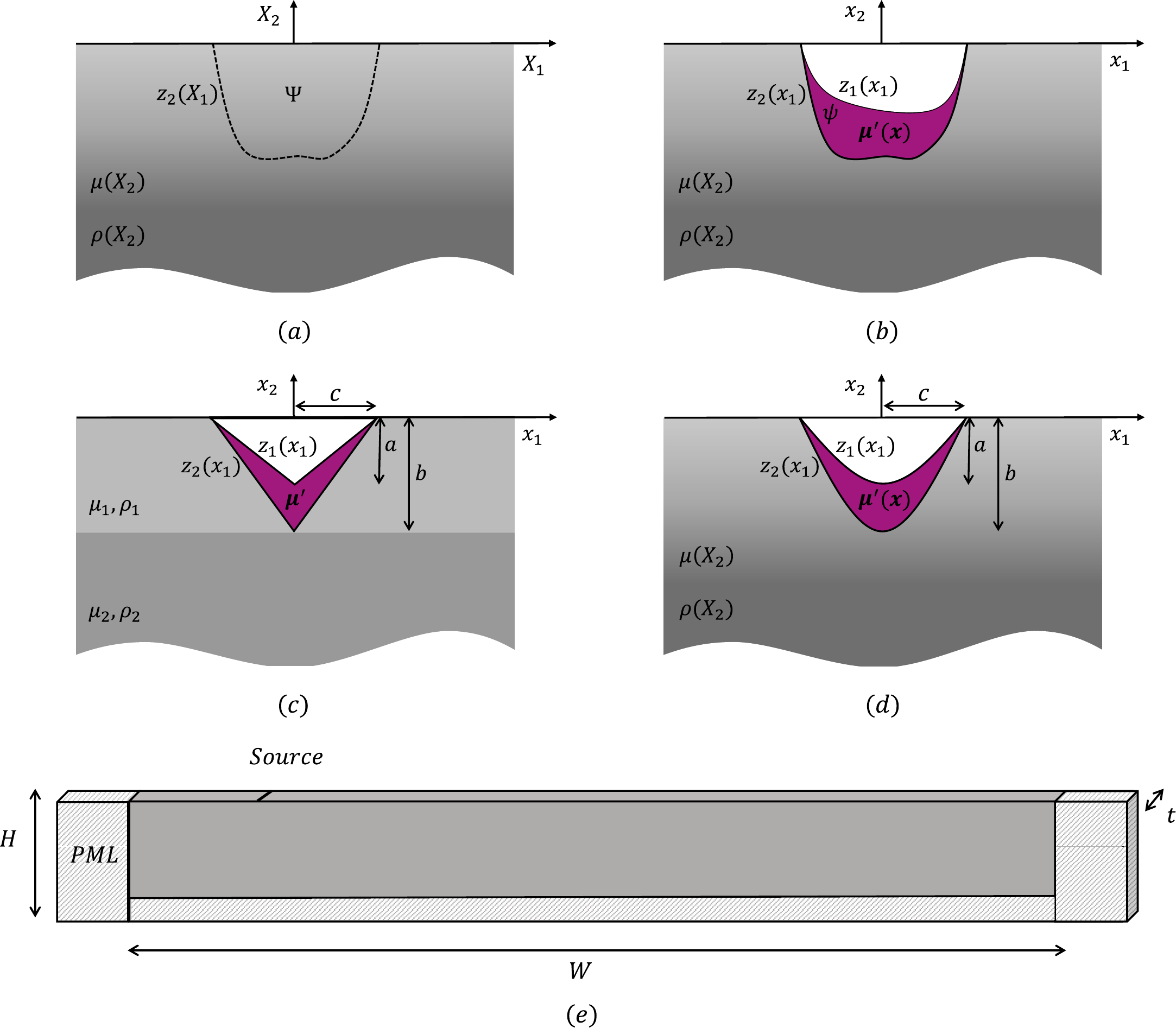}
\caption{ ($a$) Reference configuration (virtual domain). ($b$) A generic carpet-pinched cloak (purple region $\psi$) to conceal a defect (white region) from the propagating surface Love waves  (real domain). ($c$) Schematic of  a triangular pinched-carpet cloak  in a layered isotropic medium. ($d$) Schematic of a parabolic pinched-carpet cloak in an isotropic heterogeneous medium characterized by a parabolic velocity profile along the depth of the half space. ($e$) Schematic of the FE model domain used in the simulations (not in scale).}
\label{Figure 1}
\end{figure}

Following the cloaking strategy proposed by Li and Pendry in the context of electromagnetic waves \cite{li2008hiding}, we propose the design of a carpet cloak for Love waves by applying a geometric transformation at the free surface of the heterogeneous substrate. Our aim is to stretch this interface in such a way that Love waves propagate along a curved path without experiencing some scattering in the substrate.
The reference configuration consists  in an isotropic vertically heterogeneous half-space extending from  $X_2=0$, to $X_2=-\infty$ (see Fig.~\ref{Figure 1}$a$.). The  medium is characterized by shear modulus $\mu(X_2)$ and density $\rho(X_2)$ which vary along the depth of the half-space only. We consider the propagation of time harmonic antiplane shear waves travelling along the $X_1$ direction and confined at the free surface of the medium, $X_2=0$. The displacement field is restricted to the out-of-plane component $\mathbf{U}=[U_{1},U_{2},U_{3}]=[0,0,U_{3}]$ and the governing equation is given by the two-dimensional scalar equation:
\begin{equation}
\nabla_{X}\cdot\left(\mu(X_2) \nabla_{X}U_{3}{(X_1,X_2)}\right) + \omega^2\rho(X_2) U_{3}{(X_1,X_2)}=0 \label{Eq: SH waves}\end{equation}  
where $\nabla_X{=(\partial/\partial X_1,\partial/\partial X_2)^T}$ is the gradient operator with respect to the coordinates $X_1$ and $X_2$ and $\omega$ the wave angular frequency.

Here, we restrict our investigation to normally dispersive profiles, i.e., vertically heterogeneous media with increasing  shear velocity $c_s(X_2)=\sqrt{\frac{\mu(X_2)}{\rho(X_2)}}$ with depth, such that the velocity of propagation of surface waves decreases for increasing frequency \cite{foti2018}. The simplest normally dispersive configuration consists in a soft layer with shear velocity $c_{s1}$ overlying a stiffer half-space with shear velocity $c_{s2}>c_{s1}$. Antiplane surface  waves propagating in this double-layered medium are known as Love waves \cite{Love_1911}. By extension, we use the same label for antiplane surface waves propagating in any normally dispersive  half-space.

In what follows, we discuss how to hide a surface defect located in such heterogeneous, normally dispersive, half-space from the propagating Love waves.

\subsection{Geometric transformation for carpet cloaking}

Let us recall the fundamental results of transformation elastodynamic theory for an antiplane shear wave problem. Given the scalar nature of the governing equation in Eq.~\eqref{Eq: SH waves}, we can mutate the approach developed for cloaking of acoustic waves \cite{norris2008}. Hence, we consider a point-wise invertible transformation $\chi$ that maps the reference configuration (virtual domain) $\mathbf{X}$ $\in$ $\Psi$ (in Fig.~\ref{Figure 1}$a$) to the deformed region (real domain) as $\mathbf{x}=\chi(\mathbf{X}) \in \psi$ (in Fig.~\ref{Figure 1}$b$). The deformation gradients for the transformed and the reference domain are $\mathbf{F}=\nabla_{X}x$ and $\mathbf{F}^{-1}=\nabla X$, respectively. Given $\mathbf{x}=\{x_1,x_2\}$ the coordinates for the transformed domain, the forward transformation gradient $\mathbf{F}$ reads:
\begin{equation}\mathbf{F}=\nabla_{X}x=\begin{pmatrix}
\frac{\partial x_1}{\partial X_1} & \frac{\partial x_1}{\partial X_2}\\
\frac{\partial x_2}{\partial X_1} & \frac{\partial x_2}{\partial X_2}\\
\end{pmatrix}\label{Eq: F}\end{equation}
\noindent and the relation between the gradient operator in both domains is given by:
\begin{equation}\nabla_{X}=\mathbf{F}^T\nabla_{x}.
\label{Eq: Grad}
\end{equation}
By using the identity \cite{Greenleaf_2003}:
\begin{equation}
\nabla_{X}\cdot\nabla_{X}\to J\nabla_{x}\cdot(J^{-1}\mathbf{F}\mathbf{F}^T\nabla_{x}) \\
\label{Greenleaf}
\end{equation}
where $J=\rm{det}(\mathbf{F})$ is the determnant of the Jacobian matrix, we obtain the transformed governing Eq.~\eqref{Eq: SH waves} in the new coordinates $\mathbf{x}$ as:
\begin{equation}\nabla_{x}\cdot \bm{\mu}'(\mathbf{x})\nabla_{x} u_{3}(\mathbf{x})+\rho'(\mathbf{x})\omega^2u_{3}(\mathbf{x})=0 \\ 
\label{Equation 5}
\end{equation} 
with:
\begin{equation}\bm{\mu}'(\mathbf{x})=\frac{\mathbf{F}\mu(\mathbf{x}(\mathbf{X}))\mathbf{F}^T}{J}=\frac{\mathbf{F}\mu(\mathbf{x})\mathbf{F}^T}{J}, \quad \rho'(\mathbf{x})=\frac{\rho(\mathbf{x}(\mathbf{X}))}{J}=\frac{\rho(\mathbf{x})}{J}.
\label{Eq: mu rho}
\end{equation} 
Note that, in Eq.~\eqref{Eq: mu rho} the differentiation
is applied with respect to the vector variable $\mathbf{x}$ and $u_3(\mathbf{x}(\mathbf{X})) = U_3(\mathbf{X})$. The anisotropic shear modulus $\boldsymbol{\mu}'(\mathbf{x})$ and the density $\rho'(\mathbf{x})$  provided in Eq. \eqref{Eq: mu rho} describe the properties of the cloak region in $\psi$ with  $\mu(\mathbf{x}(\mathbf{X}))=\mu(\mathbf{X})$ and $\rho(\mathbf{x}(\mathbf{X}))=\rho(\mathbf{X})$ being the isotropic shear modulus and density mapped from the virtual domain $\mathbf{X}$ to the real domain $\mathbf{x}(\mathbf{X})$.
 
Let us now specify our derivation for a two-dimensional carpet-pinched cloak like the one shown in Fig.~\ref{Figure 1}b. We denote with $z_1(x_1)$ and $z_2(x_1)$ the interior and exterior boundaries of the cloak, respectively. The 
transformation $\chi$ that maps the region enclosed between two curves $(X_1; 0)$ and $(X_1; z_2(X_1))$ of the virtual domain to the one comprised between $(x_1; z_1(x_1))$ and $(x_1; z_2(x_1))$ of the real domain is:

\begin{equation}\begin{cases}
x_1=X_1\\
x_2=(1-\xi)X_2+z_1(X_1)
\end{cases}\label{Eq: map}\end{equation}
where $\xi=\frac{z_1(X_1)}{z_2(X_1)}$. Note that  $(X_1; 0)$
is mapped on $(x_1; z_1(x_1))$ while $(X_1; z_2(X_1))$ is point-wise fixed.
The  Jacobian of the transformation reads:



\begin{equation}
\mathbf{F}=\begin{bmatrix}
1 & 0\\
L & 1-\xi
\end{bmatrix}
\label{F}
\end{equation}
where $L=\frac{\partial \xi(X_1)}{\partial X_1}X_2 + \frac{\partial z_1(X_1)}{\partial X_1}$. 
By means of Eqs.~\eqref{Eq: mu rho}, we can obtain the transformed shear modulus and density within the carpet cloak  as:

\begin{equation}
\boldsymbol{\mu'}(\mathbf{x})=\begin{bmatrix}
\mu'_{11}(\mathbf{x}) & \mu'_{12}(\mathbf{x})\\
\mu'_{12}(\mathbf{x}) & \mu'_{22}(\mathbf{x})
\end{bmatrix}=\begin{bmatrix}
1 & L\\
L & (L^2+(1-\xi)^2)
\end{bmatrix}\frac{\mu(\mathbf{x})}{1-\xi}, \quad  \rho'(\mathbf{x})=\frac{\rho(\mathbf{x})}{1-\xi}
\label{Eq: mu rho carpet}
\end{equation}

Note that carpet-cloak requires a material with inhomogenous and anisotropic shear modulus and inhomogenous density. The nature of material inhomogenity depends on both the geometrical transformation and the  vertically heterogeneity of the host material properties. Conversely, the degree of anisotropicity depends only on the geometrical transformation.  In what follows, we  elucidate these aspects with two illustrative examples.
 
\section{Ideal cloaks: two illustrative examples}
\label{sec:3}

We analyse the propagation of Love waves in two distinct half-spaces: (i) a double-layered medium hosting a triangular-shaped defect (Fig.~\ref{Figure 1}$c$); (ii) a heterogeneous medium with continuous varying shear velocity profile with a parabolic-shaped defect (Fig.~\ref{Figure 1}$d$). For both configurations, pinched-carpet cloaks obtained by stretching the geometry of surface defects are considered, in accordance with the geometrical transformation in Eq.~\eqref{Eq: map}.

\subsection{Triangular pinched cloak}
\label{sec:3.1}
First, we present the simplest configuration where carpet-cloaking of Love waves can be tested, namely a triangular pinched-cloak located inside the soft layer of a double-layered semi-infinite medium. The shear velocities in the soft layer (1) and the bottom half-space (2) are $c_{s1}$ and $c_{s2}$, respectively, with $\gamma=\frac{c_{s1}}{c_{s2}}=0.3$. The density $\rho$ is assumed constant over the whole semi-infinite medium, $\rho_1=\rho_2=\rho$. We denote the depth of the soft layer by $h_1$. Numerical values of the above geometrical and mechanical quantities are collected in Table~\ref{Tab: Prop}.

We consider a triangular defect with length $2c=1.8h_1$  and depth $a=0.6h_1$. The defect is surrounded by a cloak which extends up to the depth $b=h_1$. The boundaries of the cloak are thus marked by linear functions  $z_1(X_1)=\frac{a}{c}|X_1|-a$ and $z_2(X_1)=\frac{b}{c}|X_1|-b$, where $|.|$ denotes the absolute value. The Jacobian matrix of the geometrical transformation
\begin{equation}
\mathbf{F}=\begin{bmatrix}
1 & 0\\
\sgn(x_1)\frac{a}{c} & \frac{b-a}{b}
\end{bmatrix}
\label{Eq: Jac}
\end{equation}
is constant within each symmetric half of the cloak. Similarly, the shear modulus $\mu(\mathbf{x})=\mu_1$ is constant within the cloak, since its geometry is fully contained within the  first layer. Hence, according to Eq.~\eqref{Eq: mu rho carpet},
each symmetric half of the triangular-pinched cloak presents a homogeneous and anisotropic effective shear modulus and homogeneous density given as:

\begin{equation}
\boldsymbol{\mu'}=\begin{bmatrix}
1 & \sgn(x_1)\frac{a}{c}\\
\sgn(x_1)\frac{a}{c} & \frac{(ab)^2+(bc)^2+(ca)^2-2abc^2}{(bc)^2}
\end{bmatrix}\mu_{1}\frac{b}{b-a}, \quad  \rho'= \rho\frac{b}{b-a} 
\label{Eq: mu rho triang}
\end{equation}
The  coefficients $\mu'_{11}$, $\mu'_{12}$, $\mu'_{22}$, as obtained from Eq.~\eqref{Eq: mu rho triang}, are displayed in Appendix A (Fig.~\ref{Appendix1}$(a)-(c)$).

To evaluate the cloaking capabilities of this configuration, we first perform time-harmonic simulations in a finite element environment (COMSOL Multiphysics) modeling a 3D strip of the medium with dimensions $W\times H \times t$ along the Cartesian axes $(x_1, x_2, x_3)$, embedding the described cloak (see Fig.~\ref{Figure 1}$e$). Fixed constraints are imposed at the bottom base of the model, continuity boundary conditions are imposed along the out-of-plane $x_3$ direction, and Perfectly Matched Layers are used to suppress reflections from the model terminal sections. Shear waves are generated by imposing at the surface of the model a time-harmonic line source along the $x_3$ direction located at $0.4W$ distance from the origin. The domain displacements components $u_2=u_1=0$ are restrained so to focus our investigation to antiplane waves $u_3\neq0$. 

To generalize the results of our calculations, we introduce the following normalized quantities:
\begin{itemize}

  \item $\omega^\star=\frac{\omega}{\omega_{c1}}$ the normalized circular frequency,  where   $\omega_{c1}=\frac{\pi}{h_{1}\sqrt{  \frac{1}{cs_{1}^2} - \frac{1}{cs_{2}^2}} }$ is the cut-off frequency of the first higher-order Love wave mode;
  \item $k^\star=\frac{\omega_{c1}}{c_{s2}}$, the normalized wavenumber, so that $\lambda^\star=\frac{2\pi}{k^\star}$  is the normalized  wavelength.
\end{itemize}

Accordingly, the dimensions of the model are chosen as  $W=8\lambda^\star$, $H\approx 0.69\lambda^\star$ and $t\approx 0.01\lambda^\star$.

\begin{figure}
\centering
\includegraphics[scale=0.70]{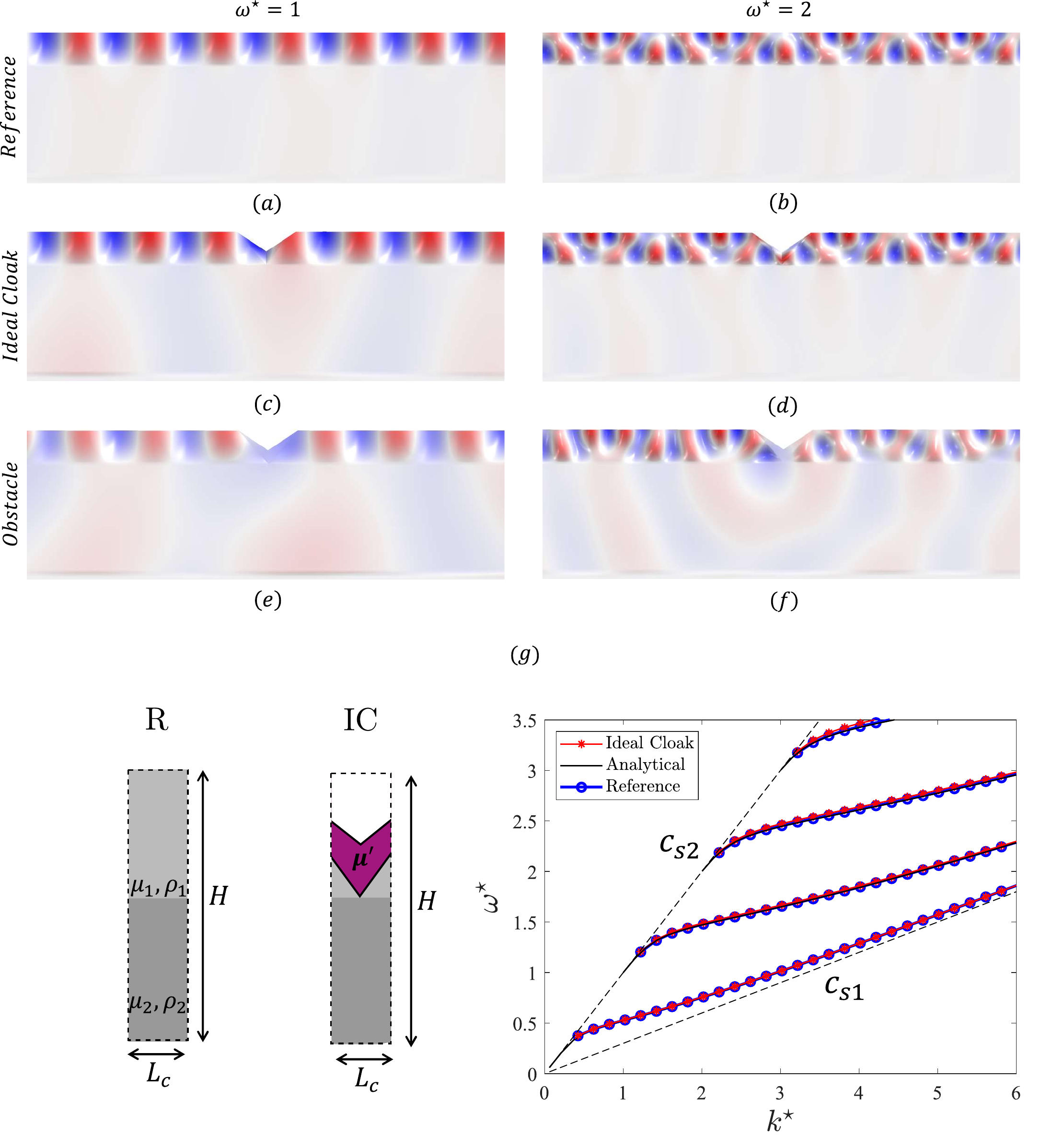}
\caption{Displacement fields of the Reference $(a)-(b)$, Ideal Cloak $(c)-(d)$ and Obstacle $(e)-(f)$ configurations, for the triangular pinched-cloak case of section 3.1 computed at frequencies $\omega^\star=1$ and $\omega^\star=2$. $(g)$ Dispersive curves (first 4 modes) for the Reference (blue lines) and the Ideal Cloak (red lines) configurations, as obtained by considering the unit cell in the schematic (R and IC), respectively. The analytical solution for the reference configuration is also presented (black lines).}  
\label{Figure 2}
\end{figure}

We compare the results of harmonic simulations performed in (i) the pristine configuration, i.e., the domain with no defect (``Reference''), (ii) in the domain including the defect (``Obstacle''), and (iii) in the domain with the cloaked defect (``Ideal Cloak''). Snapshots of the related displacements fields as obtained from FE harmonic simulations at $\omega^\star=[1, 2]$ are shown in Fig.~\ref{Figure 2}$(a)-(f)$. We remark that the propagation of Love waves has a dispersive nature: up to the first cut-off frequency the wavefield is dominated by the fundamental Love mode, see Fig.~\ref{Figure 2}$a$; conversely, for harmonic simulations at $\omega^\star=2$, Fig.~\ref{Figure 2}$b$, the total field results from the superposition of the first two modes.
Regardless the dispersive and multi-modal nature of Love waves, the triangular pinched cloak smoothly reroutes the waves around the defect (Fig.~\ref{Figure 2}$(c)-(d)$). As a result, before and after the cloaked region, the displacement field fully resemble the one of the reference configuration. The performance of the cloak are better appreciated by comparing the cloaked wave fields with those obtained in the uncloaked scenarios (see Fig.~\ref{Figure 2}$(e)$ and $(f)$). In the latter,  scattering effects can be seen both before and after the defect.

As a further proof of the broad band capabilities of the proposed carpet cloak, we compare the dispersive properties of Love waves propagating along the pristine double-layered medium and along the cloaked domain. Our aim is to prove that in the cloaked configuration, the geometrical transformation and the related transformed shear modulus in Eq. \eqref{Eq: mu rho triang} allow to fully replicate the dispersive properties of the pristine layered medium.

To this purpose, we model a portion, namely a unit cell, of the cloak in a finite element environment (COMSOL Multiphysics). The unit cell has dimension $H\times L_c \times t$ (see Fig.~\ref{Figure 2}$g$), with $L_c=\lambda^\star/20$. We apply fixed boundary conditions at the bottom surface of the cell, continuity conditions along the $x_2$ direction, and Bloch boundary conditions along the $x_1$ direction. As for the harmonic simulations, the domain displacements components $u_2=u_1=0$ are restrained. An equivalent unit cell is used to model  the ``Reference'' configuration. For all the scenarios, the  dispersive properties are obtained by solving an eigenvalue problem varying the wavenumber inside the first Brillouin zone, $k_{x_1}=[ \frac{\pi}{400L_c}; \frac{\pi}{L_c}]$ and extracting the angular frequency $\omega$. The numerical solutions are then sorted to remove spurious plate-like (and leaky) modes with phase velocity $c_p=\omega/k_{x_1}>c_{s2}$ resulting from the finite dimensions of the model. 

The dispersive curves of the ``Reference'' and ``Cloak'' configurations are reported in Fig.~\ref{Figure 2}$g$  as blue and red  lines, respectively. Markers indicate the wavenumber-frequency couples obtained from the numerical simulations. The results of the reference configuration are validated against those obtained from the solution of the classical Love wave dispersion law \cite{Love_1911} (reported in Fig.~\ref{Figure 2}$g$ as solid black lines):

\begin{equation}
\tan\left(\frac{\omega}{c_{p}}h_1\sqrt{\frac{c_{p}^2}{c_{s1}^2}-1}\right)=\frac{{\mu_2\sqrt{1-\frac{c_{p}^2}{c_{s2}^2}}}}{{\mu_1\sqrt{\frac{c_{p}^2}{c_{s1}^2}-1}}} \label{eq:12}
\end{equation}

\noindent where $c_{p}=\frac{\omega}{k}$ the phase velocity of the propagating Love waves, respectively.

The reader can appreciate how the portion of the cloaked defect supports the propagation of Love waves with dispersive properties identical to the pristine scenario. We recognize that the obtained dispersion curves describe a geometrical layout where the portion of the defect is periodically repeated along the direction of the wave propagation. 
Nonetheless, in our scenario the obtained dispersive properties are invariant w.r.t. the chosen length $L_c$ of the cell, and thus represent the effective dispersive properties of the transformed medium. More information on the dynamics of shear elastic waves across periodically perforated elastic media can be found in the recent work by Meirbekova et al. \cite{meirbekova2020}.


\begin{table}[h!]
\begin{center}
\begin{tabular}{ c|c|c|c } 
  Layer & Shear velocity & Density & Depth \\  \hline
 1 &$c_{s1}$=300 $m/s$ & 1600 $kg/m^3$ & 60 $m$  \\  \hline
 2 &$c_{s2}$=1000 $m/s$ & 1600 $kg/m^3$& $\infty$   \\  \hline
\end{tabular}
\caption{Mechanical and geometrical properties of the double-layered substrate.}
\label{Tab: Prop}
\end{center}
\end{table}

\subsection{Parabolic pinched cloak}

In the second example, we consider a vertically inhomogeneous medium characterized by a  constant density $\rho$ and shear velocity profile given by:
\begin{equation}
 c_{s}(X_2)=
\begin{cases}
 c_{s1}+\frac{c_{s2}-c_{s1}}{(h_{1})^2} (X_2)^2,  & \text{if}\quad |X_2|<|h_1|\\
  c_{s2},&\text{otherwise}

\label{Equation 12}
\end{cases}
\end{equation}
namely, a parabolic profile varying between the shear velocities $c_{s1}$ and $c_{s2}$ along the depth $h_1$ supported by a homogeneous half-space with velocity $c_{s2}$. In this medium, a parabolic-shaped defect is surrounded by a cloak enclosed within the boundaries $z_1(X_1)=\frac{a}{c^2}(X_1)^2-a$ and $z_2(X_1)=\frac{b}{c^2}(X_1)^2-b$, see Fig.~\ref{Figure 1}$(d)$.
The shear velocities $c_{s1}$, $c_{s2}$ and the geometrical parameters $a,b,c$ are identical to the ones assumed in the previous example.

The distribution of the shear modulus in the cloak region is:
\begin{equation} 
 \mu(\mathbf{x})=\rho (c_{s}(\mathbf{x}))^2
 =\rho \left( c_{s1}+\frac{c_{s2}-c_{s1}}{h_{1}^2} \left(\frac{x_2-\frac{a}{c^2}x_1^2-a}{1-\frac{a}{b}}\right) ^2 \right)^2, \quad \mathbf{x}=\chi(\mathbf{X}) \in \psi
\label{Equation 11}
\end{equation}
where  the inverse transformation of Eq.~\eqref{Eq: map} is used:
\begin{equation}\begin{cases}
X_1=x_1\\
X_2= \frac{x_2-z_1(x_1)}{1-\xi}
\end{cases}\end{equation}
The Jacobian of the transformation for the given cloak geometry is:
\begin{equation}
\mathbf{F}=\begin{bmatrix}
1 & \frac{2a}{c^2}x_1\\
\frac{2a}{c^2}x_1 & \frac{b-a}{b}
\end{bmatrix}.
\label{Eq: Jac}
\end{equation}
Overall, this leads to an anisotropic inhomogenous effective shear modulus:

\begin{equation}
\boldsymbol{\mu'}=\begin{bmatrix}
1 & \frac{2a}{c^2}x_1\\
\frac{2a}{c^2}x_1 &  \frac{4a^2}{c^4}x_1^2+\frac{(b-a)^2}{b^2} 
\end{bmatrix}\mu(\mathbf{x})\frac{b}{b-a}, \quad  \rho'= \rho\frac{b}{b-a} 
\label{Eq: mu rho parab}
\end{equation}
The reader can fully appreciate the distribution of the effective shear modulus coefficients $\mu'_{11}$, $\mu'_{12}$, $\mu'_{22}$ inside the domain of the cloak in the Figure provided in Appendix A (Fig. \ref{Appendix1}$(d)-(f)$)

\begin{figure}
\centering
\includegraphics[scale=0.68]{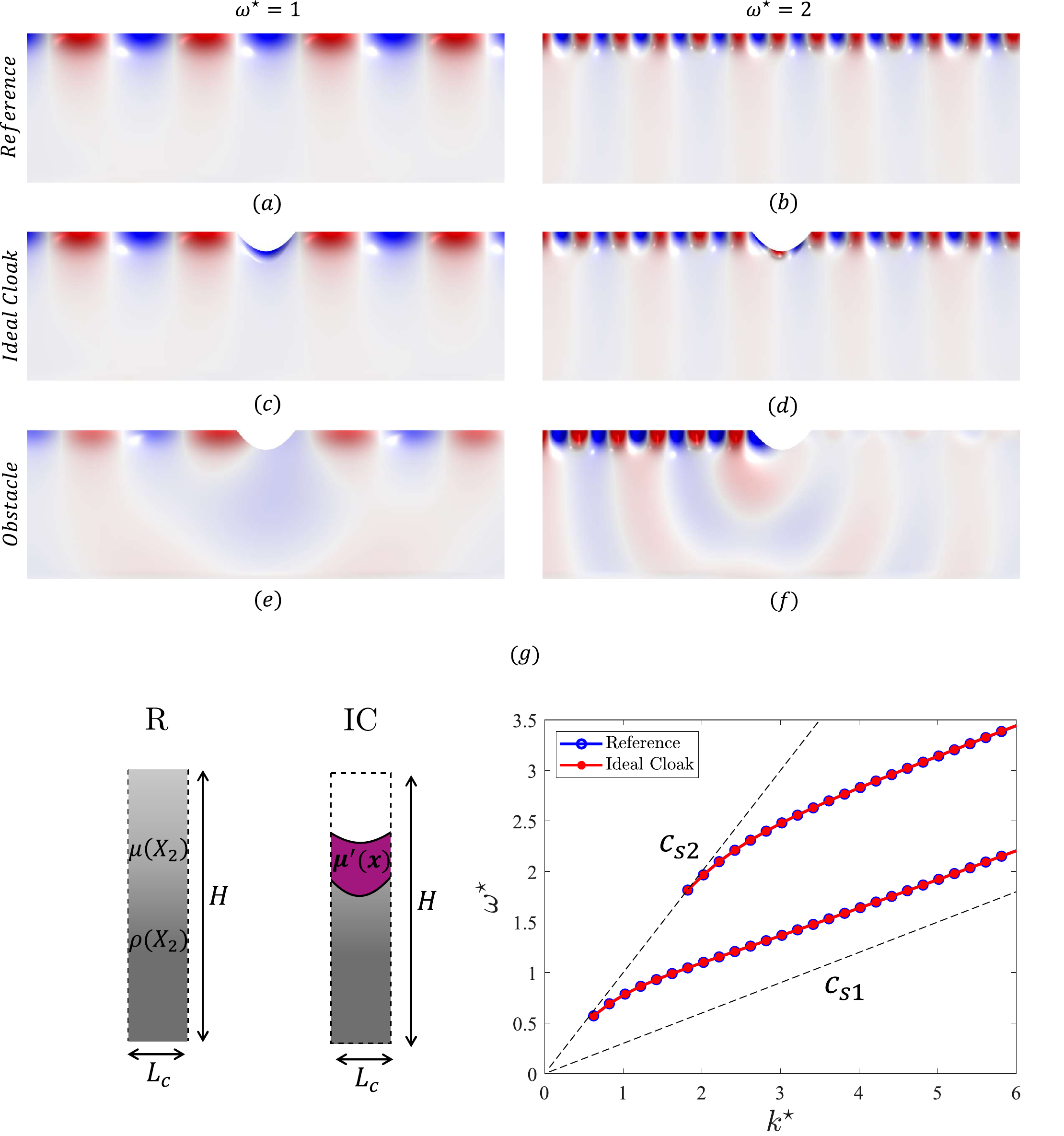}
\caption{Displacement fields for the Reference $(a)-(b)$, Ideal Cloak $(c)-(d)$ and Obstacle $(e)-(f)$ configurations, considering the parabolic pinched-cloak case of section 3.2 computed at frequencies $\omega^\star=1$ and $\omega^\star=2$. $(g)$ Dispersive curves (first 2 modes) for the Reference (blue lines) and the Ideal Cloak (red lines) configurations, as obtained by considering the unit cell in the schematic (R and IC), respectively.}  
\label{Figure 3}
\end{figure}

As for the configuration in Sect. 3.1, we test the performance of the parabolic-shaped cloak with FE harmonic simulations performed in the ``Reference'', ``Obstacle'' and ``Ideal Cloak'' scenarios. The displacements fields, obtained for harmonic sources at $\omega^\star=[1, 2]$, confirm the possibility of smoothly detouring antiplane surface waves propagating in a heterogenous medium around a defect of generic shape, see Fig.~\ref{Figure 3}$(a)-(f)$. Similarly, the dispersion analysis, conducted in analogy to what shown for the triangular cloak, corroborate the possibility of replicating the dispersive properties of the pristine scenario using the  transformed shear modulus and density obtained according to Eq.\eqref{Eq: mu rho parab},  see Fig.~\ref{Figure 3}$g$.

Although the implementation of the transformed medium in a FE numerical simulation software is straightforward, the practical realization of such material presents a number of complexities related to its heterogeneity and anisotropicity.
Thus, following a consolidated approach developed for both thermal~\cite{JI2021120948} and acoustic~\cite{Pomot_2020} cloaking, we utilize a composite medium, structured at a microscale level, to mimic at the macroscale the required mechanical properties of the transformed medium. The design of this composite material is guided by classical results of homogenization theory.

\section{Cloak realization via composite media}
We here analyse the possibility of approximating the theoretical anisotropic inhomogeneous materials obtained in the previous section by means of a layered composite material.

\subsection{Effective properties of a monoclinic layered medium}
We restrict our interest to the realization of a triangular pinched cloak.
As shown in the context of acoustics~\cite{Sun2019} and electromagnetic waves~\cite{Wang2013}, a layered periodic composite with 2D oblique geometry, i.e., a monoclinic crystal, suffices to achieve this purpose.

\begin{figure}
\centering
\includegraphics[scale=0.60]{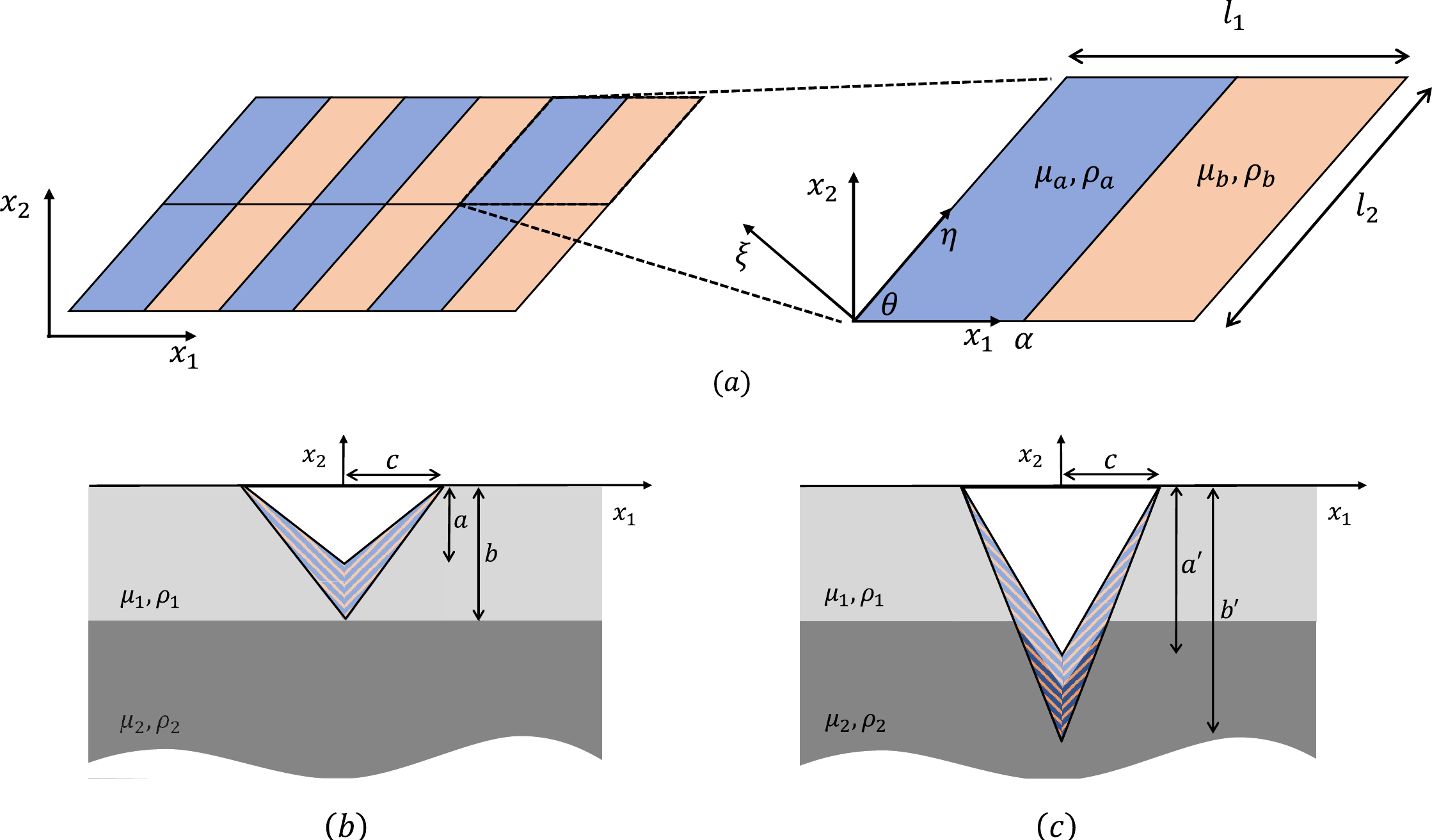}
\caption{($a$) Details of the monoclinic unit cell and ($b$-$c$) schematic representation of the layered cloaks.}
\label{Figure 4}
\end{figure}

Thus, we
consider a unit cell made of two isotropic materials, with shear moduli  $\mu_a$ and  $\mu_b$, and density $\rho_a$ and  $\rho_b$, respectively (Fig.~\ref{Figure 4}$a$). 
The lattice unit cell has dimensions $l_1\times l_2$ along the lattice vectors $e_1$ and $e_2$ which enclose the oblique angle $\theta$. The layer $a$ has a thickness $\alpha \cdot l_1$. 
Our aim is to find the geometrical parameters $\theta$ and $\alpha$ and the mechanical parameters $\mu_a$, $\mu_b$, $\rho_a$, $\rho_b$ to mimic, at the macroscale, the mechanical properties of the cloak.  
We first recall the expressions of the homogenized density  $\rho^{hom}=(1-\alpha)\rho_b+ \alpha\rho_a$ and  homogenized elastic properties of the layered cell along its principal directions $\eta$, $\xi$:

\begin{equation}
    \boldsymbol{\mu}^{hom}(\eta,\xi)=
    \begin{bmatrix}
    \langle \mu^{-1}\rangle^{-1} & 0\\
     0 &  \langle \mu\rangle\\
     \end{bmatrix}=\begin{bmatrix}
      \frac{\mu_a \mu_b}{(1-\alpha)\mu_a+ \alpha\mu_b}& 0\\
     0 &(1-\alpha)\mu_b+ \alpha\mu_a \\
     \end{bmatrix}
     \label{toto1}
\end{equation}

\noindent where $\theta$ is the  rotation angle between the principal ($\eta$, $\xi$) and reference ($x_1$, $x_2$) axes. 

Then, we diagonalize the cloak elastic tensor as:

\begin{equation}
    \boldsymbol{\mu}'(\eta,\xi)=
    \begin{bmatrix}
     \mu'_{I} & 0\\
     0 & \mu'_{II}\\
     \end{bmatrix}
     \label{diag mu}
\end{equation}

\noindent with: \begin{equation}
    \theta'=\frac{1}{2}\arctan\left(\frac{\mu'_{12}+\mu'_{21}}{\mu'_{11}-\mu'_{22}}\right)
    \label{theta}
\end{equation}
being the related rotation angle.
At this stage, we impose $\theta=\theta'$ and find the parameters $\alpha$, $\mu_a$, $\mu_b$, $\rho_a$, $\rho_b$ which fulfill the expressions:

\begin{equation}\begin{cases}
(1-\alpha)\mu_b+ \alpha\mu_a=\mu'_I\\
\frac{\mu_a \mu_b}{(1-\alpha)\mu_a+ \alpha\mu_b}=\mu'_{II}\\
(1-\alpha)\rho_b+ \alpha\rho_a=\rho'
\label{mu_I,II}
\end{cases}\end{equation}
The set of parameters which fulfill Eq.~\eqref{mu_I,II} is not unique. In the examples collected in the next section, we will fix  $\alpha$, select $\rho_a=\rho_b=\rho'$ and obtain $\mu_a$ and $\mu_b$ from the solution of Eq.~\eqref{mu_I,II}.

\subsection{Numerical examples}

We here design and show the performance of two triangular pinched cloaks, constructed using the layered medium discussed in the previous section. For the first example, we consider the geometry in  Fig.~\ref{Figure 4}$b$, where the cloak lies within the boundaries of the first layer. From Eq.~\eqref{Eq: mu rho triang}, we factor out the common term $J=\frac{b}{b-a}$ to obtain: 
 
 \begin{equation}
\boldsymbol{\mu'}=\begin{bmatrix}
1  & \sgn(x_1)0.6667\\
\sgn(x_1)0.6667 & 0.6044
\end{bmatrix}\mu_{1}, \quad \rho'=\rho
\label{mu triang small}
\end{equation}

According to Eq.~\eqref{diag mu}, the elastic tensor in the principal direction reads: 
  
\begin{equation}
\boldsymbol{\mu'}(\eta,\xi)=\begin{bmatrix}
1.4976  & 0\\
0 & 0.1068
\end{bmatrix}\mu_{1} 
\label{mu triang small}
\end{equation}
with $\theta'=\pm 0.6412$ rad.
Assuming a value of
$\alpha=0.5$, from Eq.~\eqref{mu_I,II} we obtain:
$\mu_a=0.0544\mu_1$ and $\mu_b=2.9409\mu_1$.


\begin{figure}
\centering
\includegraphics[scale=0.70]{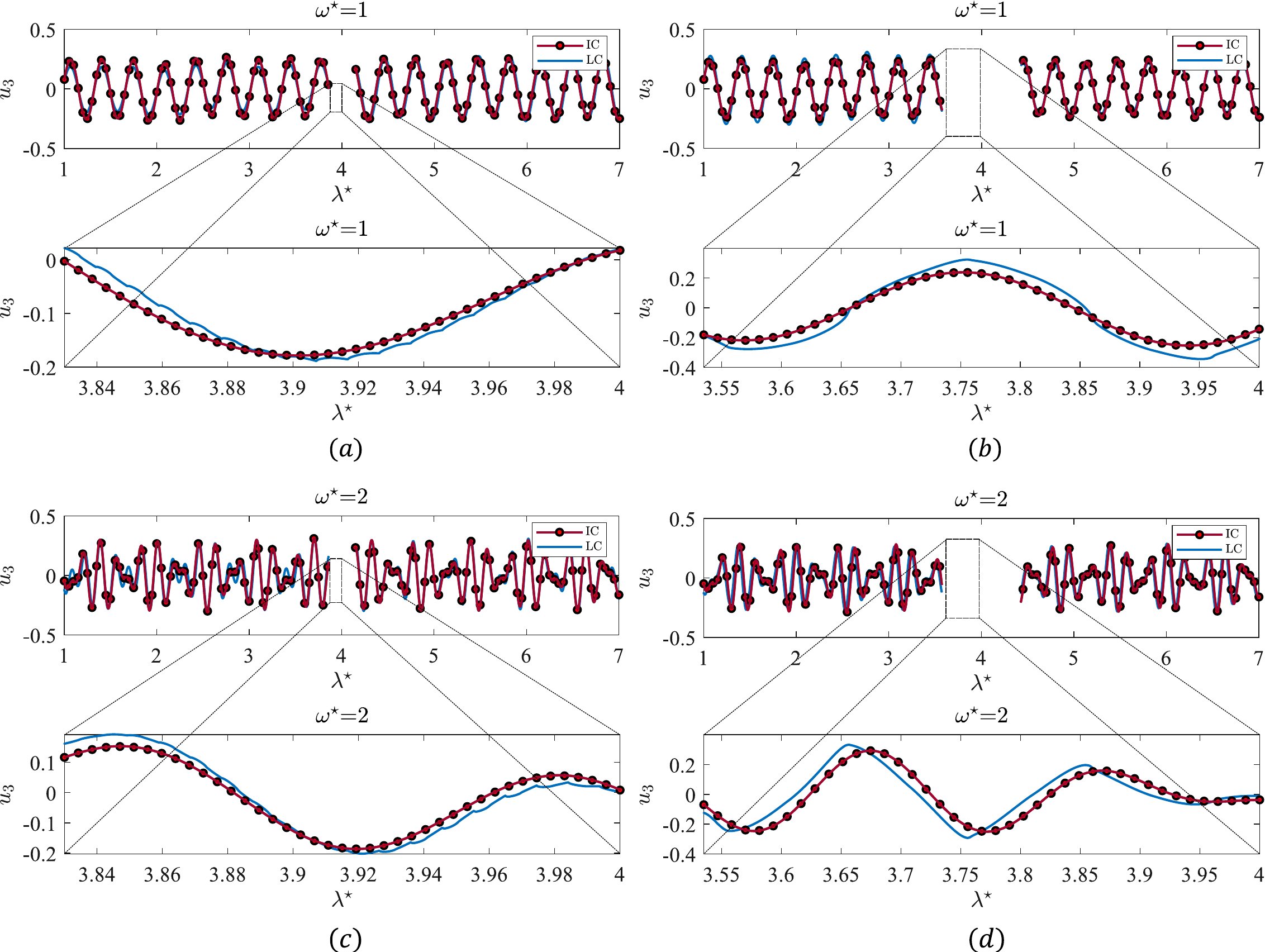}
\caption{Comparison between the Ideal (IC) and the Layered Cloak (LC) displacement fields, for a cloak fully contained inside the first layer ($1^{st}$ column, panels $(a)$ and $(c)$) and for a cloak  exceeding the first layer ($2^{nd}$ column, panels $(b)$ and $(d)$). The harmonic simulations are performed at two different normalized frequencies $\omega^\star=[1-2]$.}
\label{Figure 5}
\end{figure}

Equipped with the mechanical parameters of the layered unit cell, we perform harmonic simulation via FE models to analyse the  performance of the layer
triangular pinched cloak. The adopted FE model resembles the one used to model the ideal cloak in sect. 3.1. The layered domain comprises unit cells with dimensions $l_1=1.5$, $l_2=1.5$. As for the ideal scenario, we perform harmonic simulations at $\omega^\star=[1, 2]$

The displacement field $u_3$ extracted along the free surface of the model, $x_2=0$, before and after the cloaked region, are shown in Fig.~\ref{Figure 5}$(a)$ and $(c)$ and compared with those of an ideal cloak. The reader can appreciate how the displacement fields of the layered cloak (LC) matches the ones of the ideal case obtained from the geometric transformation. 

The same occurs inside the cloak, as confirmed by the results collected in the inset of the Fig.~\ref{Figure 5}$(a)$ and $(c)$, showing the displacement fields extracted along the boundary $z_1(x)$ of the cloak ($2^{nd}$ row in Fig.~\ref{Figure 5}$(a)$ and $(c)$). The minor discrepancies between layered and ideal cloak in Fig.~\ref{Figure 5} can be attributed to numerical errors due to the FE mesh and the finite dimensions of the unit cell. The latter lead to the presence of several partial cells at the interface between the cloak and the half-space which degrades the overall performance of the cloak. 
Such boundary layer effects, which characterize a layered medium  filled with a non integer number of cells, have been studied in \cite{maurel2018sensitivity} in the context of scalar waves. Thus, we recognize that  our effective medium formula (\ref{toto1}) would benefit from the introduction of some corrective terms depending upon the wave frequency. 
Similarly, we remark that even in the absence of such issues, improving our layered cloak's efficiency would require a very large number of cells in accordance with classical error estimates in homogenization theory, as already noticed in the context of thermal layered cloaks \cite{petiteau2014spectral}. Nonetheless,  for practical purpose, the propose design is a good trade-off between complexity and cloaking efficiency.


Following the same modelling approach, we consider a cloak with dimensions $b'$, $c$, enclosing a defect  that penetrates through the half-space (Fig.~\ref{Figure 4}$c$). For this  scenario, the layered cloak requires the use of 2 different unit cells, tiling two distinct regions of the cloak.  The properties of these unit cells are again given by $\mu_a=0.5544\mu(x_2)$ and $\mu_b=2.4004\mu(x_2)$ and $\theta'=\pm0.5544$ with:
\begin{equation}
 \mu(x_2)=
\begin{cases}
 \mu_1,  & \text{if}\quad |x_2|<|\chi({h_1})|\\
  \mu_2,&\text{otherwise}
\label{mu triang big a b}
\end{cases}
\end{equation}


Fig.~\ref{Figure 5}$(b)$ and $(d)$ ($2^{nd}$ column) shows a good agreement between the displacements for the ideal (IC) and the layered (LC) cloak along the surface.  Performance losses are again attributed to boundary layer effects since the two cloak regions are filled with a non integer number of cells. We refer once again to \cite{maurel2018sensitivity} for such issues.



\section{Conclusions}
In this work we proposed a strategy to design carpet  cloaks that make surface defects scattering free from the viewpoint of the propagation of shear polarized surface waves. Our approach relies on the invariance of the wave equation when subjected to a bi-dimensional geometric transformation of coordinates. The change of coordinates, applied to a pinched region embedding the surface defect, is used to mathematically mask the defect. Its application yields a transformed wave equation in the cloaked region which in turn  provides the local material properties to be used. Notably, in the proposed cases, the shear modulus of the material turns out to be a point dependent two-by-two fully populated
tensor and the mass density a point dependent scalar value.
We exploited these formulas to design two ideal cloaks, a triangular one and a parabolic one. Using  finite element simulations, we demonstrated the capability of the cloaks to render the surface defects transparent to Love waves. As further proof, we demonstrated that a periodically repeated portion of these cloaked defects support the propagation of classical Love waves, identical to the ones propagating in the  pristine substrate.

Finally, we provided a viable strategy to realize the triangular cloak using layered media. 
To such purpose, we used a monoclinic unit cell with two isotropic materials and exploited   results of homogenization theory to devise the dimensions of the two materials over the unit cell, their shear moduli, and the angle of the cell, such that the homogenized properties of the unit cell coincide with those derived from Eq.~\eqref{Eq: mu rho triang} for the ideal cloak. We applied this approach by shaping a second triangular cloak for a defect extending not only within the soft layer but also in the underlying half-space.  We verified, via FE simulations, the performances of the layered cloaks and found performances comparable to the ideal ones.
In terms of perspectives, experimental realizations of the proposed layered cloak  should confirm the possibility of hiding surface defects from Love waves and contribute to advancements of devices for surface waves control.
Furthermore, an analogous strategy to design carpet cloaks should be developed for Rayleigh waves. In that case, the transformed elasticity tensor loses its minor symmetry \cite{Brun2009}. Thus, the definition of an effective medium for the cloak design requires a specific treatment involving, for example, a symmetrization of the transformed elasticity tensor \cite{craster2021near}.

\section{Appendix A. Distribution of the effective shear modulus.}

\begin{figure}[h]
\centering
\includegraphics[scale=0.70]{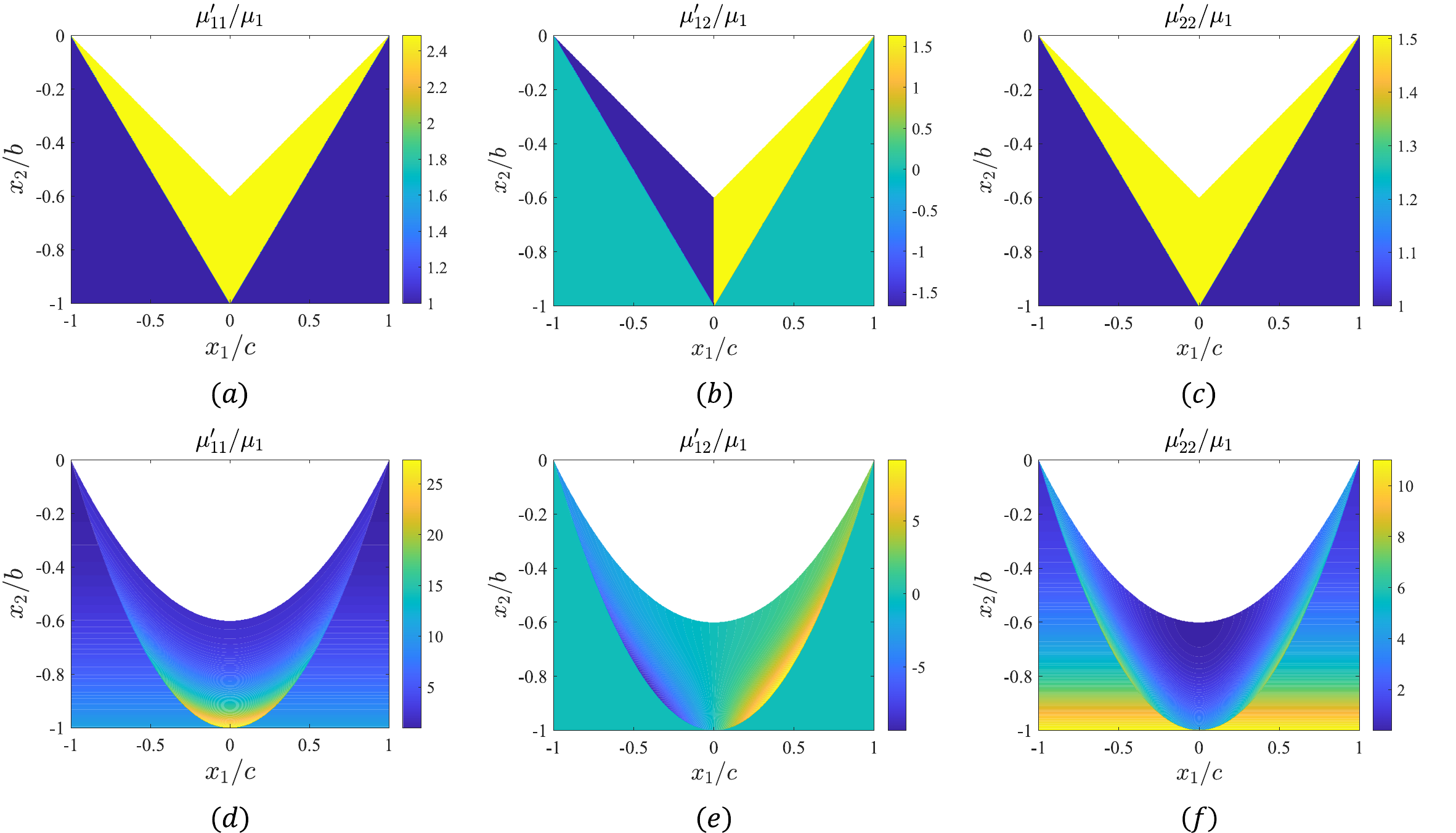}
\caption{Contour plots of the effective shear moduli coefficients over a rectangular domain $(-c,c)\times(-b,0)$, for the triangular $(a)-(c)$ and the parabolic $(d)-(f)$ shaped cloaks, respectively.}  
\label{Appendix1}
\end{figure}

In this appendix we provide the distribution of the effective shear moduli for the triangular and the parabolic cloak, as obtained from Eq.~\eqref{Eq: mu rho triang} and Eq.~\eqref{Eq: mu rho parab}, respectively. For the triangular cloak $\mu'_{11}$ , $\mu'_{12}$ and $\mu'_{22}$ are constant inside the cloak and the layer, with $\mu_{12}$ being anti-symmetric w.r.t. the cloak vertical axis (Fig.~\ref{Appendix1}$(a)-(c)$). The parabolic cloak presents a symmetric distribution for $\mu'_{11}$ and $\mu'_{22}$, and an anti-symmetric one for $\mu'_{12}$, respectively (Fig.~\ref{Appendix1}$(d)-(f)$).


\section*{CRediT authorship contribution statement}
\noindent \textbf{Zinon Chatzopoulos:} Conceptualization, Methodology, Investigation, Software, Data curation, Writing - original draft. \textbf{Antonio Palermo:} Conceptualization, Methodology, Investigation, Software, Data curation,  Validation, Writing - original draft, Co-supervision.
\textbf{Sebastien Guenneau:} Conceptualization, Methodology, Writing - review \& editing, Co-supervision.
\textbf{Alessandro Marzani:} Conceptualization, Methodology, Writing - review \& editing, Supervision, Funding acquisition.

\section*{Declaration of competing interest}
\noindent The authors declare that they have no conflict of interest.

\section*{Acknowledgments}
\noindent This project has received funding from the European Union’s Horizon 2020 research and innovation programme under the Marie Skłodowska Curie grant agreement No 813424.

\bibliography{mybibfile}

\end{document}